# The role of vegetables trade network in global epidemics

Yong Min[1], Jie Chang[1*], Xiaogang Jin[2], Yang Zhong[3], Ying Ge[1]

[1] College of Life Sciences, Zhejiang University, Hangzhou, China

[2] College of Computer Sciences, Zhejiang University, Hangzhou, China

[3] School of Life Sciences, Fudan University, Shanghai, China

**\*Corresponding Author:** College of Life Sciences, Zhejiang University, Zijingang Campus, 388 Yuhangtang Rd., Hangzhou, 310058, PR China. **Tel & Fax:** +86-571-8820 6465. **E-mail:** jchang@zju.edu.cn

**Abstract**

The outbreak of enterohemorrhagic *Escherichia coli* (EHEC) in May 2011 warns the potential threats of the world vegetables trade network (VTN) in spreading fatal infectious diseases. The heterogeneous weight distribution and multi-scale activity of intermediary networks affects the diffusion, proliferation and extinction of epidemics. Here, we constructed a dual-weighted VTN with 118 major countries and territories from FAO 2008 statistic data about global vegetation production and trade, and develop a reaction-diffusion model to simulate the epidemic behaviors in through VTN. We found an emerged asymmetric threshold of epidemic on VTN, in which local proliferation within nodes plays a more critical role than global diffusion in spreading of EHEC-like diseases, i.e. sufficient local proliferation is the precondition for global diffusion. We also found that a strong modularity on VTN structure, which restricts the spreading of EHEC-like diseases; however, within the communities, the diffusion is quick and easy. There is, moreover, a critical "epidemic stem", in which a serial of positive feedback loop for amplifying the proliferation and diffusion pathogens has been identified from entire VTN. Surprisingly, statistical analysis shows a well consistency between theoretical composition of stem and actual pattern of EHEC. The results provide a chance to design gradient control strategies for controlling disease global diffusion. Our analysis provided the first inspect of global epidemics mediated by trade networks for improved control and immunity strategies in the future.

## Introduction

There is a large body of theoretical literature on how network structures may shape the spread of infectious diseases and influence the design of optimal control strategies in past decade (Pastor-Satorras and Vespignani, 2001; Eguiluz and Klemm, 2002; Gang et al., 2005; Colizza et al., 2006; 2007;; Balcan et al., 2009; Rocha et al., 2010;). Such works usually focus on the network describing human direct interacts (Khan et al., 2009; Cauchemez et al., 2011), however, the new outbreak of haemolytic uraemic syndrome (HUS), which is a complication of enterohemorrhagic *Escherichia coli* (EHEC), poses a new challenge to current approach (Kupferschmidt, 2011a; 2011b; Dolgin, 2011). The spreading of such diseases has three new features. First, the diseases can use trade web as the medium, which includes not only a human direct interacting but also a material transportation (Frank et al., 2011). Second, the diseases involve two-scale epidemics: proliferation in a country (infected people could diffuse pathogen to environment or other agricultural products through manure to infect other people) and diffusion among worldwide (Dolgin, 2011). Third, the potential vectors of the diseases are complicated (e.g. all of cucumber, sprout or other leaf vegetables could be vectors of EHEC) (Kupferschmidt, 2011a), which lead to difficulty in identification of control targets and establishment of effective strategies. These new features and uncertainties obviously affect our ability to assess the global pattern of EHEC-like diseases and control their spreading.

Motivated by these features and challenges, here we analyze the worldwide vegetable production and trade data of 2008 from The Food and Agriculture Organization of the United Nations database (FAO, http://faostat.fao.org). Based on the dataset, the vegetable trade network (VTN) is constructed. Nodes and edges of VTN are both weighted (dual-weighted) by vegetable production and trade quantity respectively. Additionally, a dynamical reaction-diffusion model is developed to simulate two-scale epidemics in dual-weighted VTN. By simulation and theoretical analysis, we studies some general characterizations of the epidemic on VTN, including general topological properties of VTN, epidemic threshold, modularity and geographic effects, and major epidemic pathways on VTN. By appropriate integrated analysis of these characters and real epidemic pattern of EHEC, we evaluate the effectiveness of our models and, moreover, discussed the potential strategies for controlling disease global diffusion in VTN.

## Result and Discussion

**Large-Scale Structure of VTN**. FAO provides the list of 254 countries and territories (hereinafter referred to as country or countries) with the production of 574 agricultural products, and connected by trades of these agricultural products from 1986 to 2008. Because vegetables are reported as the most probable vector of EHEC in the outbreak of HUS in German, we study the FAO database and only focus on the vegetables production and trades. The VTN is a dual-weighted graph comprising 254 weighted nodes denoting countries whose weight is their vegetable production $w_i$ and 4,243 weighted edges whose weight $w_{ij}$ accounts for the direct vegetable trade between country $i$ and $j$. This data set has been compiled by remove the nodes with only inputs or outputs edges, by which these nodes cannot transmit the infectious diseases. The final network data set contains the $N$=118 nodes and $M$=3,879 edges (accounting for 99% of the worldwide vegetable production and trades). The presence of various network characters and special degree distribution in VTN indicate a possible major impact in the ensuing disease spreading pattern.

A ubiquitous character of complex networks is the so-called small-world property (Watts and Strogatz, 1998; Amaral et al., 2000). In a small-world network, pairs of nodes are connected by short paths as one expect for a random graph. In VTN, the average shortest path length $d$ is the average minimum number of direct trades that goods need to take to get from any country to any other country worldwide. We found that 56% pairs of countries are connected by two steps or less, and the average shortest path length between the 118 countries in VTN is very short, $d = 1.76$, which is shortest among all global network models (Table 1). More generally, we found that $d$ grows logarithmically with the number of countries ($n$) in the network ($d < \log N$). The $d$ of VTN is smallest among random network and all global networks, e.g. Internet or airline transportation network (Table 1) (Newman, 2003; Serrano and Boguna, 2003; Barrat et al., 2004; Colizza et al., 2006; Hu and Zhu, 2009). Additionally, some countries, which are farther in geography, are still close to each other in the network, the diameter of VTN (the largest length of all shortest pathway) is only 4. The short distance directly proofed that VTN possessed the small-world property, which could accelerate the diffusion of diseases in the networks (Kuperman and Abramson, 2001; Santos et al., 2005).

In addition to small-world, the local clustering is also bound up with the transmission behaviors on the complex networks. The clustering coefficient, $C$, quantifies the local cliquishness of a network. Here, $C$ is defined as the probability that two countries, which are directly connected to a third country, also are directly connected to each other. We find that $C$ is typically larger in VTN than in a random graph and other global networks (Table 1). These results are consistent with the expectations for a small-world network (Amaral et al., 2000), and reveal a high local cliquishness in VTN. The high local cliquishness provide a structural foundation to form a "loop" between infected countries during epidemic transmission, hence might enhance the duration and scope of diseases.

Another fundamental aspect of epidemics on complex networks is the degree distribution, that is, the distribution of the number of links of the nodes (Newman, 2003). Many real-world networks (e.g. Internet and metabolic networks) only have some nodes that are significantly more connected than others (Clauset et al., 2009). While in VTN, the probability distributions that a country $i$ has $k_i$ connections (degree of nodes) to other countries exhibit an exponential distribution, $p(k) \sim 1/\mu \cdot e^{-k/\mu}$, with very large average value, $\mu=65.75$. It is obvious different from other human epidemic networks (e.g. air-transportation network, social networks) and general worldwide trade webs (Table 1), for non-trade networks exhibit heavy-tails distribution and relatively low mean degree of each node (Table 1). Moreover, in VTN, the degrees of two nodes linked by an edge display a high negative correlation (Table 1), in which the low-degree nodes in the network associate preferentially with high-degree ones. The large average degree implies that the connection of VTN is thick, in other words, the diseases could easily spread among the countries.

All these characters, small-world, high local cliquishness and density of connections, enhance the transmission of information, material and diseases in complex networks (Newman, 2003). The fact presents an assumption that VTN could be a potential high-risk intermediate for diffusing global epidemics.

**Epidemical Dynamics on VTN**. Based on the topological structure of VTN, we develop a model based on reaction diffusion process to study the dynamic behaviors of infectious diseases spreading via dual-weighted VTN (Fig. 1). Reaction diffusion processes are used to model the phenomena as diverse as chemical reactions, population evolution, epidemic spreading and many other spatially distribution systems in which local quantities obey physical reaction equations (Colizza et al., 2007). At the microscopic level, reaction diffusion processes generally consist of particles (representing active pathogens of EHEC-like diseases) that react in the local-scale and diffuse in the global-scale. Therefore, the dynamic model could simulate EHEC-like diseases spreading on dual-weighted VTN.

In this model, for each particle A in a node, before the diffusion process, it must first react according to the basic reaction scheme,

$$A \to s A \quad (1)$$

the coefficient, $s = \{2, 3, 4, ...\}$, indicates the capacity of proliferation, i.e. one particle could proliferate into $s$ new particles. The proliferating reaction occurs with probability $\mu_i^t$, which is dependent on the weight and infected time of the node:

$$\mu_i^t = \frac{T_{max} - t}{T_{max}} \cdot \mu_i = \frac{T_{max} - t}{T_{max}} \cdot (\mu_0^{(1-\log \frac{w_i}{w_{max}})}) \quad (2)$$

where $T_{max}$ is the maximal epidemic duration of nodes and represent the immunization speed within countries, $w_{max}$ is the maximal weight of nodes in the network, and $\mu_0$ ($0 \geq \mu_0 \geq 1$) is the basic reacting probability. The particles, which do not react, will be removed and not participate in next diffusion process.

After proliferating reaction, each obtained particle in node $i$ could transmit to the direct neighbors of the node with probability indicated by the weights of both node and edge:

$$\beta_{i,j} = \frac{w_{i,j}}{\sum_k w_{k,j}} \beta_0^{(1-\log \frac{\sum_k w_{k,j}}{w_i + \sum_k w_{k,j}})} \quad (3)$$

where $\beta_0$ ($0 \geq \beta_0 \geq 1$) is the basic diffusing probability. The particles without diffusion are reserved for reacting in next time step.

In sum, there are four input parameters: the reacting coefficient $s$, maximal epidemic duration $T_{max}$, basic reacting probability $\mu_0$, and basic diffusing probability $\beta_0$. For the epidemic process, $s$ and $\mu_0$ define the infectious behaviors within a country, and $T_{max}$, indicates the speed of immunization (the number of steps that has been infected could maintain for a country), and $\mu_0$ define the diffusion behaviors among worldwide. The model is specially being designed for potential epidemics (e.g. EHEC) in VTN for integrating two features. First, the reaction diffusion model is analogous to the classic epidemic models (e.g. SIS and SIR model) and holds on classical properties (e.g. phase transition and threshold). Second, the model utilizes both weight of node and edge in VTN to simulate two-scale epidemical activities: proliferation within a country and diffusion among the countries. In the following, the results shown refer to this specific epidemic model.

**Threshold of Epidemics on VTN**. Global epidemic would be extremely relevant in the case of the emergence of a new disease that in general spreads rapidly with substantial transmission occurring before the onset of case-defining symptoms (Pastor-Satorras and Vespignani, 2001; Eguiluz and Klemm, 2002; Colizza et al., 2007). In epidemic on complex networks, the existence and value of nonzero thresholds plays the critical role in study epidemic dynamics, hence we determined the critical value of $\mu_0$ and $\beta_0$ for fix value of $s$ and $T_{max}$, . We carried out $10^4$ runs for each value of $\mu_0$ and $\beta_0$ to plot threshold line of three network models, including VTN and random networks (see *Materials and Methods*). The shape and relative position of threshold curves in Fig. 2 a showed the diffusion of epidemics is difficult in VTN than in random networks. The minimal $\mu_0$ and $\beta_0$ for global occupying all nodes on VTN is 8~10 times larger than two types of random networks. Moreover, $s$ and $T_{max}$ also exists critical value for global epidemic in VTN (Fig. 2 b and c). When setting $\mu_0 = \beta_0 = 1$, we found that $s$ and $T_{max}$ should be larger than 5 and 7 respectively. The dynamical simulating result is completely contrary to the assumption presented in previous section of structural analysis. VTN easily spread diseases in topology, however, the heterogeneous distribution of node and edge weights, which follow the power-law distribution with coefficient $Y = 1.7~1.8$, will dramatically suppress the diffusion on VTN. The result supported previous theoretical studies that epidemic spreads more quickly on unweighted networks than on weighted networks with the same condition (Barrat et al., 2004; Gang et al., 2005).

According to the previous studies, different probabilities play a balance role in determining the epidemical behaviors (Pastor-Satorras and Vespignani, 2001; Colizza et al., 2007). Surprisingly, the critical value of $\mu_0$ is obviously larger than critical value of $\beta_0$ (Fig. 2 a). When $\beta_0 = 1$, the minimal value of $\mu_0$ leading to global epidemic still over 0.8, contrarily, the minimal value of $\beta_0$ is only 0.1 when $\mu_0 = 1$. These asymmetric thresholds occurred in VTN suggests that the local proliferation within nodes plays a critical role in global spreading of EHEC-like diseases. Moreover, in contrast with previous studies indicated that the global epidemic behavior is governed by network structure on worldwide scale (Balcan et al., 2009); our dual-weighted model reveals the importance of local behavior for epidemic on networks. The reason can be attributed to the unbalance of heterogeneous distribution of node and edge weights. Although both the two types of weight have heavy tails, the distribution of node weights is more disperse than the distribution of edge weights. For example, China accounts for nearly half of global vegetable production, but the distribution of trade is relatively centralized; whose variance (5.64e$^{-4}$) is only ~1/200 of production (8.72e$^{-3}$).

The asymmetric thresholds in VTN also illustrated that local proliferation plays a critical role in global infection, which provide an inspiration in control strategies for decision makers. For epidemics mediated by VTN, decision makers should pay more attention on control infectious behaviors within countries rather than restrict trade between countries. Such strategies help to reduce the international trade friction, e.g. the argument between German and Spain on this EHEC outbreak.

**Effects of Community and Geography**. Community structure, i.e. groups of nodes that have a high density of connectivity within them and a lower density of connectivity between them, is another important feature of complex networks could affect the transmission of infectious diseases (Variano et al., 2004; Salathe and Jones, 2010). We detect community structure in VTN by modularity maximization algorithm developed by Newman (2004; 2006). The algorithm is rigorous because it maximizes an explicit parameter $Q$, identifies the number of non-overlapping communities, assigns membership to communities, and tests the significance of the results. Usually modularity *Q* > 0.3 states the existence of communities in the target networks. On feeding VTN through the algorithm, we discover that the network divides cleanly into six non-overlapping communities and, remarkably, the maximal modularity reached *Q*=0.492. In Fig.3, we illustrated the resulted six communities of VTN, and found, interestingly, the division of communities in VTN highly identify with the geographic division, which is agreed with previous results (Guimera et al., 2005). Six identified communities could be geographically described as Central Asia and Russia, Pacific and Indian Rim, Europe and North Africa, South America, North and Central America, and Southern Africa. In fact, 82.67% of vegetable trade occurs within the six communities. Except of asymmetric thresholds mentioned above, the consistency of community and geographic distribution also suggested that the epidemic on VTN is also restricted by the geographic effects.

Even though geographical distance plays a clear role in the definition of the communities of VTN, the composition of some of the communities cannot be explained solely by geographical consideration. The community that contains China displays zonal distribution along the coast of the Pacific and Indian Ocean, and across three continents and 26 countries. Additionally, from the perspective of geography, the Africa is the most complex. Although the geographic distance is restively close to each other, the countries in Central and North Africa belong to three different communities with countries in other continents, and the countries of South Africa form an independent community. These exceptions suggested that geographical distance is not unique factor in determining the trade between countries. Therefore, network community is more accurate in reflecting the trade relationship than geographical distance, and can be use to make more effective strategies in control the spreading of diseases on VTN.

Modularity prevents the infinite spreading of diseases in the complex networks (Variano et al., 2004). It suggests that the global spreading of infectious diseases in VTN is difficult; however, the spread within the communities displays a different situation (). For all six communities, the thresholds of global occupation for communities are lower than the entire VTN and even homogeneous weighted random networks (Fig. 3). Interestingly, the real epidemic pattern of EHEC support the theoretical result, where 94% (15 of all 16 countries) belong to one community identified from VTN. The results suggested that infectious diseases mediated by VTN tend to be a regional infection rather than global infection. Therefore, in addition to independent country, the network community is the next-level critical targets, to which decision makers should be pay attention for effectively controlling the epidemics.

**Global Rank of Countries.** In order to characterize the role of each country in the VTN base on its pattern of transmission disease, we first distinguished the nodes that play the role of proliferation and diffusion. Note that countries are powerful in proliferating pathogens, but they are not effective in spreading the pathogens to other countries due to relative lower exports, like China. According to the equation 3, large production will lead to a "dilution effect", i.e. for given volume, the larger amount of production; the smaller is the chance that any pathogen will be diffused. Contrarily, although the capability of proliferation is not remarkable, countries still have strong potential to spread the diseases, like Netherland. Thus, an integrated measure is necessary to illustrate the role of countries by combining the measure of proliferation and diffusion.

The betweenness is a centrality measure of a node or edge within a network and is an indicator of its importance to epidemic dynamics (Puzis et al., 2007; Bobashev et al., 2008; Meloni et al., 2009). The betweenness here is defined as the number of shortest paths connecting any two nodes that involve a transfer at a node or edge. However, classical algorithm of shortest paths cannot directly transfer to our dual-weighted network; hence a mathematical definition of "distance" between two nodes should be provided firstly. We assume an epidemical process starting from a node *i*, with one pathogen particle, where initiative node cannot be infected by other nodes again. For each edge, $e_{ij}$, we can get three generating functions:

$$f_t(x) = \mu_i^t x^s + (1-\mu_i^t) \quad (4)$$

$$g(x) = (1-\beta_{i,j})x \quad (5)$$

$$h(x) = \left(\frac{1-\sum_j \beta_{i,j}}{1-\beta_{i,j}}\right)x + \left(\frac{\sum_j \beta_{i,j} - \beta_{i,j}}{1-\beta_{i,j}}\right) \quad (6)$$

where, the $f(x)$ define the reacting probability distribution of node $i$ at time $t$, $g(x)$ indicate the probability that a particle do not diffuse through edge $e_{ij}$, and $h(x)$ is the conditional probability distribution that a particle in node $i$ do not diffuse through edge $e_{ij}$ but go through other edges. These generating functions defined the life cycle of particles within a node; thus the nest of three generating function could tell us the probability of utilization of edge $e_{ij}$:

$$U_{i,j} = 1 - H_{T_{max}} \quad (7)$$

where $H_{t+1}(x) = H_t(f_t(g(h(x))))$ ( $0 \le t \le T_{max}$ ), $U_{ij}$ is the function of $s$, $T_{max}$, $\mu_0$, and $\beta_0$, hence it could integrate the capability of proliferation and diffusion and represent the distance between two nodes.

Based on $U_{ij}$, we can calculate betweenness of nodes ($B^n$) and edges ($B^e$), and obtain the rank of countries for their importance of spreading diseases. Surprisingly, the rank is highly consistent with the real epidemical pattern of EHEC (Table 2). The average of betweenness of 16 countries reported HUS or EHEC is 437.38, which is about 14 times larger than average betweenness of other countries (Table 2). When compared by the Kolomogoroy-Smimov test, the difference between the reported countries and other countries showed that two group were significantly different from each other (p < 0.001). Such result proofed that our VTN and reaction diffusion model is effective in study spread of EHEC-like diseases, and would predict potential pattern of global infection for establish more accurate control strategies.

**The Epidemic Stem of VTN**. The explicit epidemic pattern could be more useful in controlling diseases than rank the countries. In order to quantify the potential pattern of EHEC-like diseases spreading in VTN, we adopt an additional measure, $R$, of the sharpness of the boundary of sub-network to classify the nodes and edges into meaningful sub-networks, i.e. the "stem" of disease spreading.

According to the rank of node betweenness, we removed low-betweenness nodes one by one, and plot the curve of $R$ for sub-network of rest nodes with all relative edges. The two inflection points of the curve (Fig.4 a) divided all nodes into three different roles: (1) "ultra-core nodes", i.e. nodes with betweenness $B > 348$; (2) "core nodes", $B > 12$; (3) "peripheral nodes", $B < 10$. The ultra-core and core nodes cover 50% reported countries, and account for 98% cases of HUS and EHEC in real epidemic dataset (Table 2). We found, the ultra-core and core nodes are the critical countries in different communities, and the connection between these nodes could be the major "bridge" of global epidemic among different communities.

In order to explore the connecting pattern of ultra-core and core node groups, we perform the same method as above by removing low-betweenness edges.. There also two inflection points for each node groups in classifying edges (Fig.4 b and c). When we select minimal interconnected (i.e. there is no any isolated node which do not connect to any other nodes in node groups) set of edges for each node group (22 edges for ultra-core group and 146 for core group), the ultra-core and core node combining with classified edges form the "stem" of VTN in spreading disease (Fig. 4 d). Interestingly, the core nodes rarely interact with each other, but almost entirely connect with the ultra-core nodes. Contrarily, by performing $10^4$ times simulation, we found that 40% transmitting activities occur within the stem, which only account for 24% of total nodes and 4% of total edges (refs). The result illustrated that there are many infectious loop within ultra-core nodes (coexistence of edges from node $i$ to node $j$ and reversely), and form a positive feedback style to amplify the proliferation and diffusion of diseases.

The stem, in fact, play a critical role in connecting different communities at worldwide through vegetable trade. The connection style of stem presents a radial shape from ultra-core nodes (most of nodes belong to Europe community) to core nodes (including all six communities) (Fig. 4 d). Therefore, the stem combining with single country and community to form a three-level regulation targets in control diseases worldwide.

## Conclusion

We carried out a "systems" analysis of the structure and dynamic of the VTN. The study enables us to unveil a number of significant results. The VTN is a small-world, high-density and high-cluster network in topology, but asymmetric threshold, high modularity and geographic effect could suppress diffusion of diseases. Our analysis of structure and dynamic of VTN is important for two additional reasons. First, it provides a gradient control strategy, which makes a practical trade-off between cost and efficiency in controlling disease global diffusion. The existence of asymmetric threshold, community and stem suggested that control relatively narrow collection of countries could get well effect in suppressing EHEC-like diseases. Second, although all but 5 of HUS and EHEC cases were in people who had travelled to or lived in Germany during the incubation period of infection (World Health Organization, http://www.euro.who.int/), unexpectedly, the analysis of trade network could provide us a pattern that is very close to real situation. The fact could be explained by the highly relationship between trade and travel (Kulendran and Wilson, 2000; Fischer, 2004; Wong and Tang, 2010), hence VTN not only can model trade behaviors, but also can be a potential model for studying global epidemics. Third, although current cases of HUS and EHEC are attributed to trail among countries, the legacy effect of trade and production (i.e. the proliferation via agricultural production and the diffusion via trade usually cost more time than human direct interaction via travel) could lead to a potential threaten in long-term, thus the VTN should be valued in infectious diseases. It is the first time that trade networks are considered as a medium of global epidemics, and our study also offers a quantitative and general approach to understand the two-scale epidemic behavior on dual-weighted network model.

## Materials and Methods

**Monte Carlo Simulation**. We performed Monte Carlo simulations of reaction diffusion model on VTN and random networks: (1) dual-weighted VTN; (2) heterogeneous random networks with the same mean degree and weights of VTN; and (3) unweighted random networks with the same mean degree of VTN. The topology of random networks is generated from both configuration model with exponential degree distribution and Watts-Strogatz (WS) model. These network models process the similar topological structures (i.e. degree distribution), but the major difference of these models is attributed to the weights distribution. In each Monte Carlo step, we randomly perform reaction and diffusion according to the probability defined in equation 2 and 3. Initial conditions are constructed in randomly choosing a node as the source of diseases and assigning one particle to the node. Transients are discarded and simulations run long enough that the initial conditions do not affect the results.

**Threshold Detection Algorithm**. In order to detect the transition of whether reaching the global occupied and find threshold of transition in this situation, we perform a binary search algorithm for basic diffusing probability $\beta_0$ for given basic reacting probability $\mu_0$ at within domain [$a$, $b$], where $a = 0$ and $b = 1$ initially.

Step 1: If $b-a<E$, we find the threshold, $(b-a)/2$, and the algorithm terminates, where $E$ is the minimal error value allowed by the algorithm, and $E = 0.01$ in this paper.

Step 2: Otherwise, we repeat $10^4$ times Monte Carlo simulation to test whether occur global occupied when $\beta_0 = (b-a)/2$.

Step 3: If occupied, $b = (b-a)/2$, otherwise $a = (b-a)/2$, and repeat Step 1.

For each given $\mu_0$ and calculated $\beta_0$, we can plot the threshold lines in Fig. 2 a.

## Acknowledgements

## Reference

Amaral LAN, Scala A, Barthélémy M, Stanley HE (2000) Classes of small-world networks. *Proc. Natl. Acad. Sci. USA* 97:11149 -11152.

Balcan D et al. (2009) Multiscale mobility networks and the spatial spreading of infectious diseases. *Proc. Natl. Acad. Sci. USA* 106:21484 -21489.

Barrat A, Barthélemy M, Pastor-Satorras R, Vespignani A (2004) The architecture of complex weighted networks. *Proc. Natl. Acad. Sci. USA* 101:3747-3752.

Bobashev G, Morris RJ, Goedecke DM (2008) Sampling for Global Epidemic Models and the Topology of an International Airport Network. *PLoS ONE* 3:e3154.

Cauchemez S et al. (2011) Role of social networks in shaping disease transmission during a community outbreak of 2009 H1N1 pandemic influenza. *Proc. Natl. Acad. Sci. USA* 108:2825-2830.

Clauset A, Shalizi CR, Newman MEJ (2009) Power-Law Distributions in Empirical Data. *SIAM Rev.* 51:661.

Colizza V, Barrat A, Barthélemy M, Vespignani A (2006) The role of the airline transportation network in the prediction and predictability of global epidemics. *Proc. Natl. Acad. Sci. USA* 103:2015-2020.

Colizza V, Pastor-Satorras R, Vespignani A (2007) Reaction-diffusion processes and metapopulation models in heterogeneous networks. *Nat Phys* 3:276-282.

Dolgin E (2011) As E. coli continues to claim lives, new approaches offer hope. *Nat Med* 17:755.

Eguíluz VM, Klemm K (2002) Epidemic Threshold in Structured Scale-Free Networks. *Phys. Rev. Lett.* 89:108701.

Fischer C (2004) The influence of immigration and international tourism on the demand for imported food products. *Food Economics - Acta Agriculturae Scandinavica, Section C* 1:21-33.

Frank C et al. (2011) Epidemic Profile of Shiga-Toxin–Producing *Escherichia coli* O104:H4 Outbreak in Germany — Preliminary Report. *N Engl J Med*:110630110615077.

Gang Y, Tao Z, Jie W, Zhong-Qian F, Bing-Hong W (2005) Epidemic Spread in Weighted Scale-Free Networks. *Chinese Phys. Lett.* 22:510-513.

Hu Y, Zhu D (2009) Empirical analysis of the worldwide maritime transportation network. *Phys. A* 388. Available at: http://www.sciencedirect.com/science/article/pii/S0378437108010273 [Accessed September 14, 2011].

Khan K et al. (2009) Spread of a Novel Influenza A (H1N1) Virus via Global Airline Transportation. *N Engl J Med* 361:212-214.

Kulendran N, Wilson K (2000) Is there a relationship between international trade and international travel? *Appl. Econ.* 32:1001-1009.

Kuperman M, Abramson G (2001) Small World Effect in an Epidemiological Model. *Phys. Rev. Lett.* 86:2909.

Kupferschmidt K (2011) Scientists Rush to Study Genome of Lethal E. coli. *Science* 332:1249-1250.

Kupferschmidt K (2011) As E. coli Outbreak Recedes, New Questions Come to the Fore. *Science* 333:27.

Meloni S, Arenas A, Moreno Y (2009) Traffic-driven epidemic spreading in finite-size scale-free networks. *Proc. Natl. Acad. Sci. USA* 106:16897-16902.

Newman MEJ (2003) The structure and function of complex networks. *SIAM Rev.* 45:167--256.

Newman MEJ (2006) Modularity and community structure in networks. *Proc. Natl. Acad. Sci. USA* 103:8577-8582.

Newman MEJ, Girvan M (2004) Finding and evaluating community structure in networks. *Phys. Rev. E* 69:026113.

Pastor-Satorras R, Vespignani A (2001) Epidemic Spreading in Scale-Free Networks. *Phys. Rev. Lett.* 86:3200.

Puzis R, Elovici Y, Dolev S (2007) Fast algorithm for successive computation of group betweenness centrality. *Phys. Rev. E* 76:056709.

Rocha LEC, Liljeros F, Holme P (2010) Information dynamics shape the sexual networks of Internet-mediated prostitution. *Proc. Natl. Acad. Sci. USA* 107:5706-5711.

Salathé M, Jones JH (2010) Dynamics and Control of Diseases in Networks with Community Structure. *PLoS Comput Biol* 6:e1000736.

Santos FC, Rodrigues JF, Pacheco JM (2005) Epidemic spreading and cooperation dynamics on homogeneous small-world networks. *Phys. Rev. E* 72:056128.

Serrano MÁ, Boguñá M (2003) Topology of the world trade web. *Phys. Rev. E* 68:015101.

Variano EA, McCoy JH, Lipson H (2004) Networks, Dynamics, and Modularity. *Phys. Rev. Lett.* 92:188701.

Watts DJ, Strogatz SH (1998) Collective dynamics of /`small-world/' networks. *Nature* 393:440-442.

Wong KN, Tang TC (2010) Tourism and openness to trade in Singapore: evidence using aggregate and country-level data. *Tourism Economics* 16:965-980.

**Tables**

Table 1. The compare of global complex networks.

|  | type | $n$ | $m$ | Distribution | $<k>$ | $<d>$ | $C$ | $R$ |
|---|---|---|---|---|---|---|---|---|
| VTN | directed | 118 | 3,879 | $p_k \sim e^{-k/65.75}$ | 65.75 | **1.76** | **0.66** | -0.25 |
| WTW | directed | 179 | 7,510 | $p_k \sim k^{-2.6}$ | 83.91 | 1.80 | 0.65 | - |
| Internet | undirected | 10,697 | 31,992 | $p_k \sim k^{-2.5}$ | 5.98 | 3.31 | 0.39 | -0.19 |
| WWW | directed | 269,504 | 1,497,135 | $p_k \sim k^{-2.1}$ | 5.55 | 11.27 | 0.29 | -0.07 |
| ATN | directed | 3,880 | 18,810 | $p_k \sim k^{-2.0}$ | 9.70 | 4.73 | 0.62 | + |
| WMN | directed | 878 | 7,955 | $p_k \sim k^{[-1.7\ -2.9]}$ | 18.12 | 3.60 | 0.40 | ? |

Table 2. The structure properties of reported countries from WHO[*].

|  | HUS | EHEC | Betweenness | Degree |
|---|---|---|---|---|
| Austria | 1 | 4 | 0 | 99 |
| Canada | - | 1 | 109 | 153 |
| Czech Republic | - | 1 | 0 | 82 |
| Denmark | 10 | 15 | 0 | 102 |
| France | 7 | 10 | 510 | 176 |
| Germany | 857 | 3,078 | 357 | 178 |
| Greece | - | 1 | 4 | 107 |
| Luxembourg | 1 | 1 | 0 | 24 |
| Netherlands | 4 | 7 | 3,109 | 185 |
| Norway | - | 1 | 0 | 87 |
| Poland | 2 | 1 | 38 | 113 |
| Spain | 1 | 1 | 381 | 162 |
| Sweden | 18 | 35 | 5 | 106 |
| Switzerland | - | 5 | 109 | 121 |
| United Kingdom | 3 | 4 | 348 | 172 |
| United States of America | 4 | 2 | 2,028 | 194 |

The data of HUS and EHEC come from WHO (World Health Organization, http://www.euro.who.int/).

**Figure Legends**

**Fig. 1. The reaction diffusion model in networks**. Schematic representation of reaction diffusion in complex networks. Particles can diffuse in the network and, inside each node, undergo the reaction process described by equation (1).

**Fig. 2. Phase diagram and power-law distribution of network weights**. (a) Phase transition line for basic reaction and diffusion probability when $s = 6$ and $T_{max} = 5$. Five different network models are plotted. (b) Phase transition line for reaction capacity, $s$. (c) Phase transition line for maximal epidemic duration, $T_{max}$. (d) The power-law distribution of edge weights (i.e. trade quantities) with coefficient -1.72. (e) The power-law distribution of node weights (i.e. product quantities) with coefficient -1.79. (d) and (e) are plotted and fitted by the tools from Clauset et al., 2009.

**Fig. 3. Communities in the worldwide vegetable trade network**. Each color corresponds to a community. The six line arts represent the phase transition lines for basic reaction and diffusion probability when $s = 6$ and $T_{max} = 5$ on six communities.

**Fig. 4. The definition of epidemic stem of vegetable trade network**. (a) The local modularity line with two inflection points as removing low-betweenness nodes to classify the nodes into ultra-core, core and peripheral groups. (b) The local modularity line with two inflection points to classify the edges of core group. (c) The local modularity line with one inflection point to classify the edges of ultra-core group. (d) The structure of the epidemic stem, the inner cycle represent the ultra-core group, and outside cycle is core group. The colors of node represent the communities, to which node is belong.

**Figures**
**Fig. 1.**

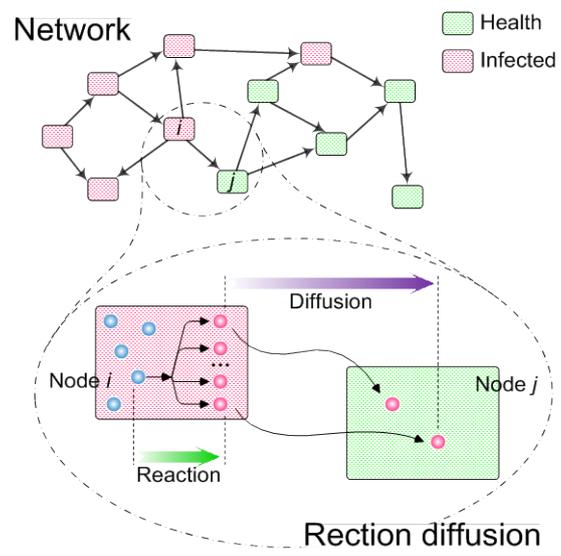

**Fig. 2.**

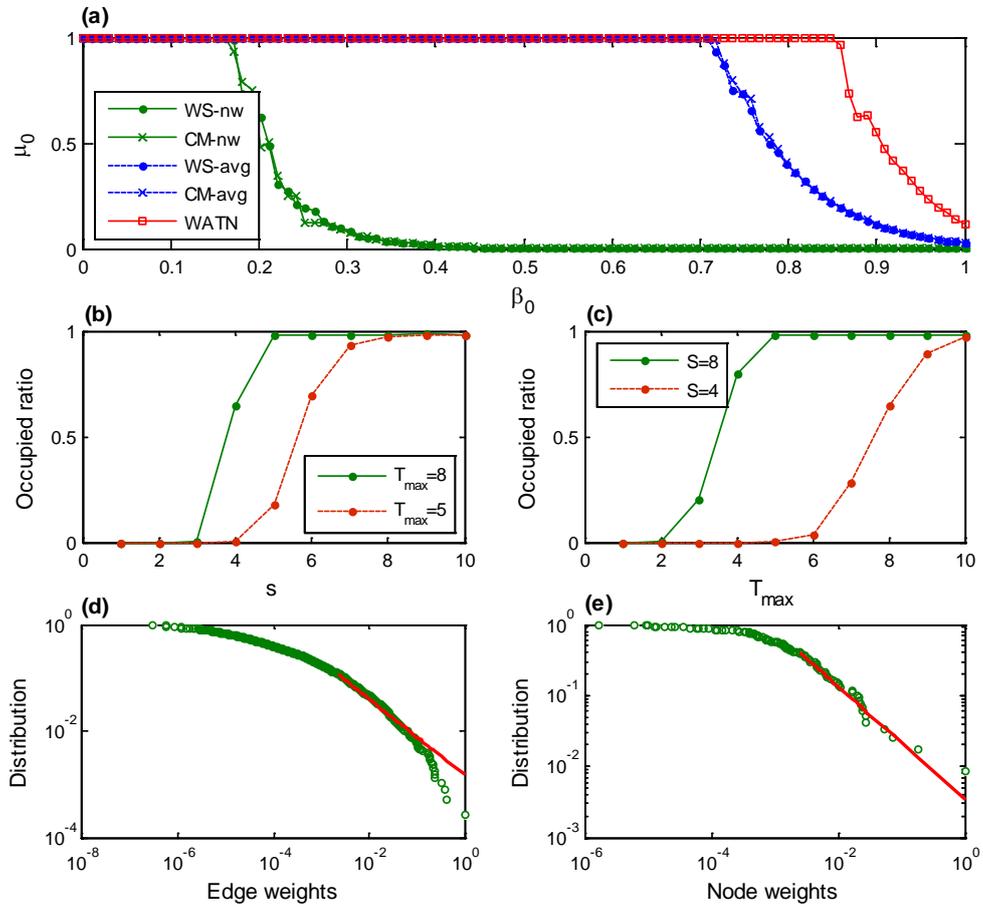

**Fig. 3.**

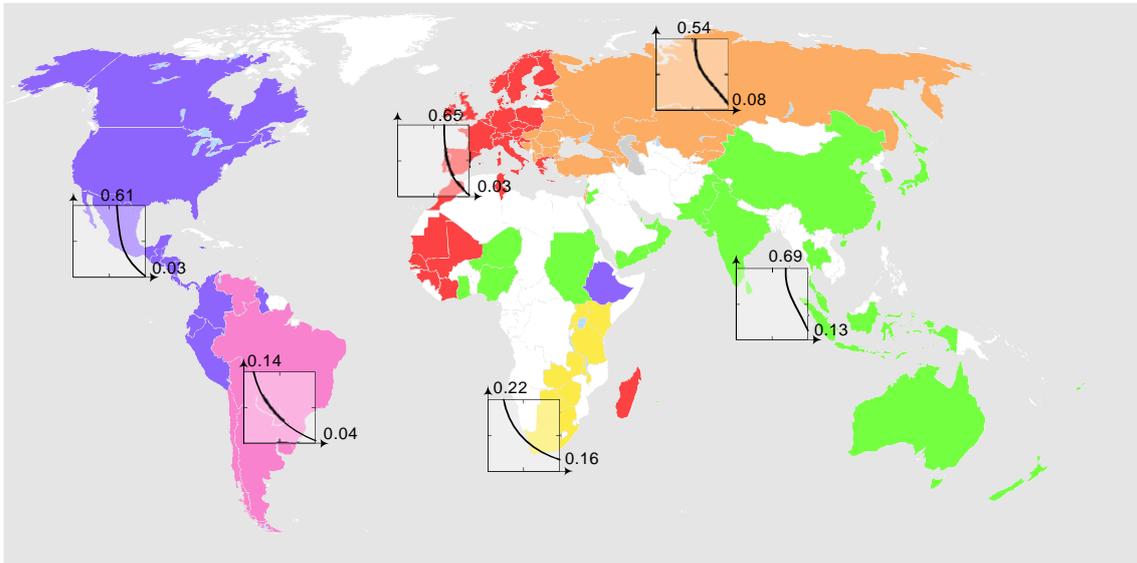

**Fig. 4.**